\documentclass[online]{aa}
\usepackage[varg]{txfonts}
\RequirePackage{hyperref}
\hypersetup{
	colorlinks = true,
	linkcolor = blue,
	citecolor = blue,
	filecolor = blue,
	urlcolor = blue,
	pdfborder = {0 0 0}
}
\newcommand{\doiurl}[1]{\href{https://doi.org/#1}{#1}}
\newcounter{IonCS}
\DeclareRobustCommand{\ion}[2]{\setcounter{IonCS}{#2}\textup{#1\,\large{\scshape{\roman{IonCS}}}}}
\graphicspath{{./}{Bilder/}}

\newcommand{\pan}[1]{\textbf{#1)}}
\newcommand{\sect}[1]{Sect.\,\ref{S:#1}}
\newcommand{\sects}[2]{Sect.\,\ref{S:#1} and \ref{S:#2}}
\newcommand{\fig}[1]{Fig.\,\ref{F:#1}}

\newcommand{\graphflex}[4][figure]{\begin{#1}#2\caption{#4}\label{F:#3}\end{#1}}

\newcommand{\graphwidthflex}[6][figure*]{\graphflex[#1]{#5\includegraphics[width=#4]{#2.pdf}}{#3}{#6}}
\newcommand{\graphwidth}[4][15cm]{\graphwidthflex{#2}{#3}{#1}{\centering}{#4}}

\newcommand{\graphside}[4][12.8cm]{\graphwidthflex{#2}{#3}{#1}{\sidecaption}{#4}}

\newcommand{\eqi}[1]{$#1$}

\DeclareRobustCommand*{\unit}[1]{~\ensuremath{\mathrm{#1}}}
\renewcommand{\AA}{\r{A}}

\RequirePackage{color}
\definecolor{darkgreen}{rgb}{0,0.45,0}

\newcommand\blfootnote[1]{%
	\begingroup
	\renewcommand\thefootnote{}\footnote{#1}%
	\addtocounter{footnote}{-1}%
	\endgroup
}

\begin{document}

\setlength{\topmargin}{-20pt}
\AANum{A123}
\yearCop{2013}
\doi{\doiurl{10.1051/0004-6361/201321185}}
\idline{A\&A 555, A123 (2013)}
\hypersetup{
	pdftitle = {Observationally driven 3D~MHD model of the solar corona above an active region},
	pdfauthor = {Ph.-A.~Bourdin, S.~Bingert, and H.~Peter},
	pdfkeywords = {Sun: corona -- magnetohydrodynamics (MHD) -- methods: numerical -- Sun: UV radiation},
	pdfsubject = {Astronomy \& Astrophysics}
}

%
\title{Observationally driven 3D~MHD model\\of the solar corona above an active region$^\star$}
\titlerunning{Observationally driven 3D~MHD model of an AR corona}

\author{Ph.-A.~Bourdin\inst{1,2}, S.~Bingert\inst{1}, and H.~Peter\inst{1}}
\authorrunning{Ph.-A.~Bourdin et al.}

\institute{%
Max-Planck-Institut f{\"u}r Sonnensystemforschung, 37191 Katlenburg-Lindau, Germany
\\
\email{Bourdin@MPS.mpg.de}
\and
Institut f{\"u}r Astrophysik, Universit{\"a}t G{\"o}ttingen, Friedrich-Hund-Platz 1, 37077 G{\"o}ttingen, Germany
\vspace{6pt}
}

\date{Received 29 January 2013 / Accepted 9 May 2013}

\abstract%
{%
}
{%
The goal is to employ a 3D~magnetohydrodynamics (MHD) model including spectral synthesis to model the corona in an observed solar active region.
This will allow us to judge the merits of the coronal heating mechanism built into the 3D~model.
}
{%
Photospheric observations of the magnetic field and horizontal velocities in an active region are used to drive our coronal simulation from the bottom.
The currents induced by this heat the corona through Ohmic dissipation.
Heat conduction redistributes the energy that is lost in the end through optically thin radiation.
Based on the MHD model, we synthesized profiles of coronal emission lines which can be directly compared to actual coronal observations of the very same active region.
}
{%
In the synthesized model data we find hot coronal loops which host siphon flows or which expand and lose mass through draining.
These synthesized loops are at the same location as and show similar dynamics in terms of Doppler shifts to the observed structures.
This match is shown through a comparison with Hinode data as well as with 3D~stereoscopic reconstructions of data from STEREO.
}
{%
The considerable match to the actual observations shows that the field-line braiding mechanism leading to the energy input in our corona provides the proper distribution of heat input in space and time.
From this we conclude that in an active region the field-line braiding is the dominant heating process, at least at the spatial scales available to current observations.
}
\keywords{ Sun: corona -- magnetohydrodynamics (MHD) -- methods: numerical -- Sun: UV radiation \vspace{6pt}}

\maketitle

\section{Introduction\label{S:intro}}

Many processes have been identified that are able to deliver a sufficient amount of energy at the base of the corona to heat the plasma to more than \eqi{10^6\unit{K}} \citep[e.g.,][]{Klimchuk:2006,McIntosh+al:2011,Wedemeyer-Boehm+al:2012}.
One of them is Ohmic dissipation of currents that are induced by the braiding of magnetic field lines rooted in the photosphere \citep{Parker:1972}, which we use in this work.
Recently, \cite{Cirtain+al:2013} claimed to have directly observed this braiding.
The goal of the present study is to investigate the coronal structure and dynamics resulting from this process by means of a forward model.
We synthesize emission line profiles from a numerical 3D~magnetohydrodynamics (MHD) model that allows a direct comparison to actual observations.
This provides a crucial test for the distribution of the heat input in space and time through the field-line braiding process.

Previous studies modeled the global magnetic structure of the Sun and reproduced actual observations with a \emph{prescribed} coronal heating function \citep[e.g.,][]{Lionello+al:2005}.
The first proper implementation of Parker's field-line braiding process was achieved by \cite{Gudiksen+Nordlund:2002,Gudiksen+Nordlund:2005a,Gudiksen+Nordlund:2005b}.
To fit the active region into the computational box, they had to downscale the domains side-length by a factor of five. This reduces the total magnetic flux and, more importantly, this eliminates the magnetic-field patches of the network by averaging in space.
\blfootnote{$^\star$ Parameters and simulation log-files are only available at the CDS via anonymous ftp to \href{http://cdsarc.u-strasbg.fr}{\texttt{cdsarc.u-strasbg.fr}} (\href{ftp://130.79.128.5}{\texttt{130.79.128.5}}) or via \mbox{\url{http://cdsarc.u-strasbg.fr/viz-bin/qcat?J/A+A/555/A123}}}
Thus, the magnetic connections from the core of the active region to the surrounding network are not included.
Nonetheless, with this model \cite{Gudiksen+Nordlund:2002,Gudiksen+Nordlund:2005a,Gudiksen+Nordlund:2005b} found a loop-dominated corona, where synthesized emission line profiles reproduced observations in a statistical sense \citep{Peter+al:2004,Peter+al:2006}, in particular concerning the persistent redshifts in the transition region \citep{Peter+Judge:1999,Peter:1999full}.
In these and in later studies (c.f. \sect{coronal.model}) the comparison to observations is done statistically or by comparing typical structures.
These models were not compared directly with observations by matching the magnetic field at the lower boundary in the photosphere and at the same time reproducing the observed coronal emission.

\graphwidth[15.75cm]{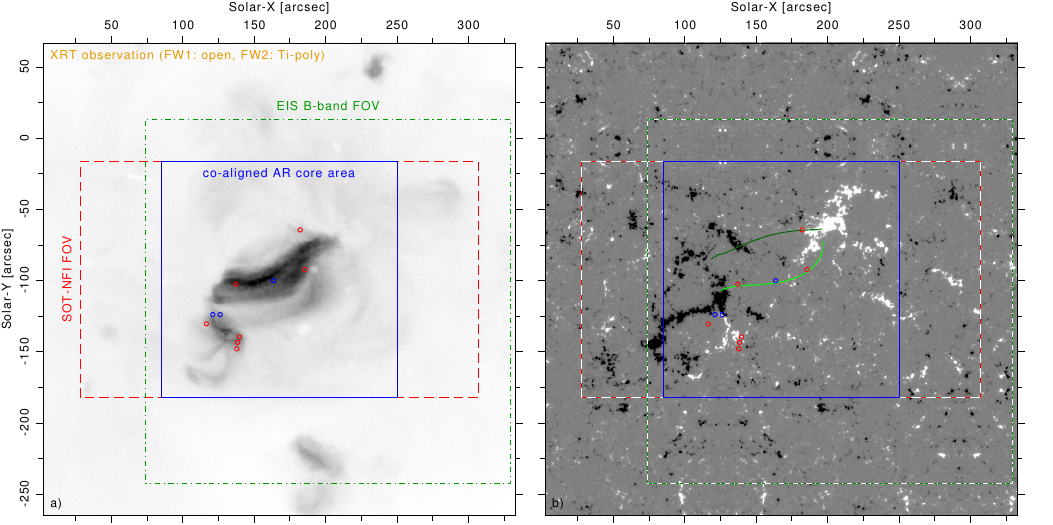}{overview}{
Active region observed by the Hinode satellite on 14\,Nov\,2007.
The left panel shows the X-ray emission observed by XRT together with the field-of-view of EIS (green dash-dotted square) and SOT-NFI (red dashed rectangle).
The right panel displays a line-of-sight magnetogram (saturation level: \eqi{\pm 300\unit{G}}) of the active region that is smoothly embedded in a quiet Sun carpet (see \sect{boundary.condition}).
We use only the co-aligned AR core area (blue solid square) for our analyses.
The circles and lines indicating various loop structures are co-spatial with those in \fig{comparison}.}

In this new study we aim at a one-to-one comparison between a 3D~MHD model and observations to test the field-line braiding mechanism.
For the first time we have done this for a \emph{full active region at the correct spatial scale} in a large numerical experiment matching the observable spatial resolution.
Therefore, this is a major step towards a realistic description of the corona in a 3D~model.

While the 3D~model cannot resolve the actual dissipation length scales that go down to the meter scale and below, it does provide a self-consistent treatment of the energy input, redistribution, and radiative losses to get a proper coronal energy balance.
This redistribution of energy, in particular the heat conduction along the magnetic field, is essential to self-consistently set the coronal plasma pressure, which is a prerequisite when synthesizing coronal emission that is to be compared to actual observations.
Because of the limitations of the spatial resolution our model as well as previous 3D~MHD models do not resolve the individual nanoflare reconnection events proposed by \cite{Parker:1988}.
The process actually described in the numerical models might be better characterized as \emph{magnetic diffusion}.

We first discuss the general model strategy (\sect{strategy}) before giving some details on the model setup (\sect{model}) and presenting our results (\sects{hot.loops}{reconstruction}).

\section{Model strategy\label{S:strategy}}

The central idea behind this study is to compare synthesized emission from a forward 3D~MHD coronal model driven by photospheric observations to actual coronal observations.
For this we use observations of the magnetic field and horizontal velocities in the photosphere to prescribe the lower boundary of the 3D~MHD model.
From the model we synthesize emission line spectra which are observable with current extreme UV spectrographs, and are thus directly comparable to coronal observations.

For our study we select an active region (AR) for which observations have been taken simultaneously in the photosphere and in the corona (\fig{overview}).
We use a data set from the Hinode solar space observatory \citep{Kosugi+al:2007}, which includes observations from the X-ray telescope XRT, spectra of \ion{Fe}{12} and \ion{Fe}{15} from the extreme UV imaging spectrometer \citep[EIS, ][]{Culhane+al:2007}, and the spectro-polarimeter (SP) and narrowband filter imager (NFI) of the solar optical telescope \citep[SOT, ][]{Tsuneta+al:2008}.
The SP and the NFI provide vector- and line-of-sight magnetograms, and horizontal velocities in the photosphere.

The active region under investigation did not show sunspots, but a set of hot loops is visible in X-rays (\fig{overview}a).
These connect two extended regions of strong magnetic field with opposite polarity (\fig{overview}b).
We use a time series of magnetograms to define the magnetic field and horizontal velocities at the bottom boundary of the computational domain (see \sect{boundary.condition}).

Data from Hinode/EIS provide a raster map of the active region, including the \ion{Fe}{12} (195\unit{\AA}) and \ion{Fe}{15} (284\unit{\AA}) emission lines.
From these we derive the intensity and Doppler shifts (\fig{comparison}a,b).
After a careful spatial alignment (\sect{alignment}) we can then compare the coronal observations to the synthetic model data (\sect{hot.loops}).

The coronal model is powered by the observed photospheric magnetic field that is advected by the observed photospheric horizontal velocities.
This leads to field-line braiding and induces currents in the corona that are dissipated and heat the plasma.
The 3D~MHD model provides the temperature, density, and velocity at each grid point of the computational domain.
Following the approach of \cite{Peter+al:2004,Peter+al:2006} we use the atomic data base {\sc Chianti} \citep{Dere+al:1997,Young+al:2003} to synthesize emission lines.
This provides maps of intensity and Doppler shift that can be compared directly to the coronal observations.

This strategy enables us to test our model and the underlying theoretical assumptions for the coronal energy input, i.e., braiding of magnetic field lines and the subsequent Ohmic dissipation of induced currents.
The aim is to check if the model description is sufficient to reproduce realistic coronal structures and their dynamics.

\section{3D~MHD model and alignment with observations\label{S:model}}
\subsection{Coronal model\label{S:coronal.model}}

\graphside[12.7cm]{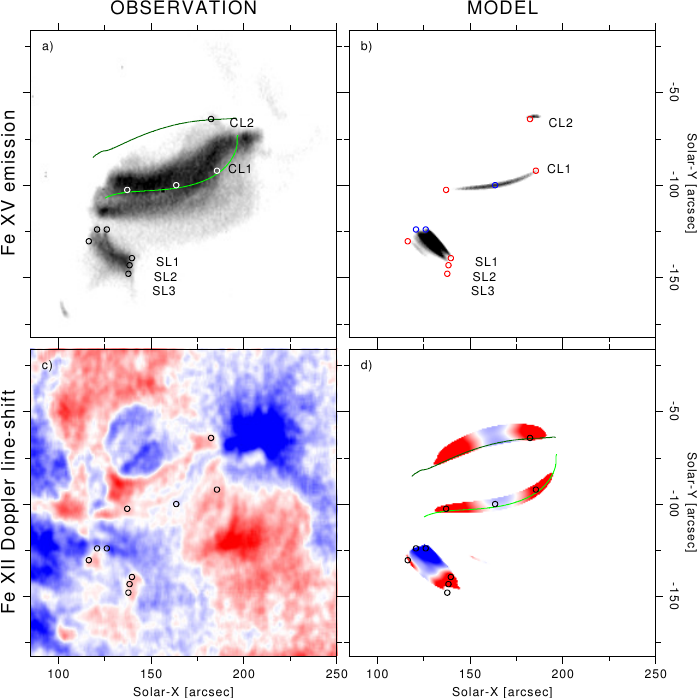}{comparison}{
Direct comparison between observations and the 3D~MHD forward model.
The left column shows the active region as observed by EIS on 14\,Nov.\,2007.
Panel \pan{a} displays the intensity map in \ion{Fe}{15} (284\unit{\AA}) on a linear inverse scale normalized to the peak intensity. For plotting we set a threshold of 1/6 of the peak intensity, which is well above the noise level.
In panel \pan{c} we plot the Doppler map in \ion{Fe}{12} (195\unit{\AA}).
The scale of the doppler map covers \eqi{\pm 10\unit{km/s}}, where blue-shifts indicate plasma flows towards the observer.
The right column shows the corresponding quantities synthesized from the 3D~MHD model with the same color coding.
A short loop system can be seen spanning from one of the two main polarities to the network of the quiet Sun (SL\,1--3), as well as a loop (system) in the core of the AR between the two main polarities (CL\,1).
In the model we traced two magnetic field lines, rooted in the centers of CL\,1 and 2, that are overplotted in green.
The circles are located at the same positions in all panels.
The alignment between the observations and synthesized images is accurate within about 3\unit{arcsec} corresponding to the diameter of the circles.\vspace{-2pt}}

The basic setup of our numerical experiments follows the philosophy of \cite{Gudiksen+Nordlund:2002,Gudiksen+Nordlund:2005a,Gudiksen+Nordlund:2005b} and \cite{Bingert+Peter:2011}.
The initial condition of our model consists of a stratified atmosphere in hydrostatic equilibrium.
We initialize the magnetic field configuration with a potential field extrapolation from the observed photospheric magnetogram.

For the temporal evolution of the model we employ the compressible resistive MHD equations to compute the temperature, velocity, density, and the magnetic vector potential inside the computational domain.
The photospheric driving advects the magnetic field, which induces currents \eqi{j} in the upper atmosphere that lead to Ohmic heating \eqi{\mu_0 \eta j^2}.
For the magnetic diffusivity \eqi{\eta} and also for the kinematic viscosity we use a constant value of \eqi{10^10\unit{m^2/s}} in the corona.
For details of the model the reader is referred to \cite{Bingert+Peter:2011}.

As outlined by \cite{Bingert+Peter:2011} we choose \eqi{\eta} so that the current sheets that form have a finite width still resolved by our numerical scheme; in other words, we choose \eqi{\eta} so that the magnetic Reynolds number is of the order of unity when choosing the grid spacing as a length scale.
Therefore our heating term \eqi{\mu_0 \eta j^2} should be considered a parameterization of the true heating mechanism.
A full model, including the actual dissipation process \emph{and} covering a macroscopic structure observable on the Sun (e.g. a whole active region) is beyond current capabilities.
Some steps of models in the context of solar flares going beyond the MHD picture including kinetic processes, for example, can be found in \cite{Cargill+al:2012}.

In contrast to our approach, \cite{Lionello+al:2005} use a \emph{prescribed} heating function in their 3D~MHD model. There magnetic energy is dissipated through an explicit term in the equations or by numerical diffusion, but this is not consistent with the chosen heating function.
In our approach, the magnetic energy actually dissipated (in the induction equation) is converted into heat (through \eqi{\mu_0 \eta j^2} in the energy equation).
We consider our treatment to be more consistent than simply prescribing a heating function.

Our model includes gravity, heat conduction parallel to the magnetic field following \cite{Spitzer:1962}, and optically thin radiative losses based on \cite{Cook+al:1989}.
The heat conduction is of pivotal importance because it sets the pressure in the corona, and thus is essential if one wants to compare the synthesized emission to actual observations.

Summing up, we have a self-consistent description of the thermal structure of the plasma and of the magnetic field in the coronal structures.
With a spatial resolution of down to 100\unit{km} this 3D~MHD model cannot match the resolution possible in 1D loop models, of course.
Still, the implementation of the field-aligned heat conduction together with the optically thin radiative losses allows us to properly describe the energy cycle between the chromosphere and the corona.
Here the Ohmic heating of the corona leads to heat conduction back to the chromosphere, which together with the local heat input there leads to evaporation of material that then expands into the corona.
The description of this cycle is important because it basically sets the pressure of the coronal structure \citep{Withbroe:1988}.
With the limits of the spatial resolution in a 3D~model the temperature gradients are less steep than in a 1D loop model, but still the energy cycle between the chromosphere and corona is accounted for.

Models similar to the one presented here were able to produce a loop-dominated corona \citep{Gudiksen+Nordlund:2002,Gudiksen+Nordlund:2005a,Gudiksen+Nordlund:2005b} where synthesized average quantities matched observables such as the differential emission measure and the transition region Doppler shifts \citep{Peter+al:2004,Peter+al:2006,Hansteen+al:2010,Zacharias+al:2011a}.
Furthermore, these models provided a new way to understand loops with constant cross section \citep{Peter+Bingert:2012} and provided insight in the spatio-temporal distribution of the heat input into the corona \citep{Bingert+Peter:2011,Bingert+Peter:2013}.

To run the numerical experiments, we use the Pencil Code \citep{Brandenburg+Dobler:2002}\footnote{\url{http://Pencil-Code.Nordita.org/}}.
The parameters of the simulation are available at the Centre de Donn\'ees astronomiques de Strasbourg (CDS).
The computational domain covers \eqi{235 \times 235\unit{Mm^2}} horizontally and 156\unit{Mm} vertically with \eqi{1024 \times 1024 \times 256} grid points.
The horizontal grid spacing is 230\unit{km}, which is roughly the spatial resolution of the magnetograms employed for the photospheric driving of our model.
In the vertical direction we use a non-equidistant grid to resolve the strong gradients in temperature and density with a resolution of about 100\unit{km} up to the transition region.

We advanced the model in total for about 65\unit{min} solar time.
After about 50\unit{min}, the model reaches a state independent of its initial condition.
During the last 15\unit{min}, strong Ohmic heating sets in and the peak temperature in the box rises from 0.5\unit{MK} to about 1.4\unit{MK}, where the rapid increase comes to a halt.
The system reaches a quasi-stationary state and individual structures develop.

\subsection{Lower boundary condition from observations\label{S:boundary.condition}}

The observed magnetogram time-series has a cadence of 90\unit{seconds}.
From that we deduce horizontal motions of the magnetic patches by local correlation tracking.
The typical spatial scale of these patches is 15\unit{Mm} (about 10~granules) and their velocity distribution peaks at 100\unit{m/s}.
With this method the solar granulation on a scale of 1\unit{Mm} remains unresolved.
Therefore, we generate a horizontal velocity field by using a method described in \citep{Gudiksen+Nordlund:2002} that matches statistical properties of observed granulation and that we have used before \citep[e.g.,][]{Bingert+Peter:2011}.
The velocity field we use as a driver in our simulation is the superposition of the observed flow field and the generated field on smaller scales.

The field-of-view of the NFI data covers just the active region magnetic field concentration (see \fig{overview}b).
Therefore, it is not large enough for our simulation, because this would cause problems with the side boundary conditions, which are in our case periodic.
Therefore we smoothly embed the observed AR inside a periodic carpet of mirrored quiet Sun (QS) magnetogram patches that we also took from observations (see \fig{overview}b).
In this process we ensure that the magnetogram at the bottom is periodic.
This additional QS area isolates the main magnetic patches in the periodic setup.
This ensures a more realistic magnetic field topology and allows field lines to connect from the main polarities into the QS network.
For the calibration of the magnetograms we use several snapshots of the AR core that are available as spectro-polarimetric SOT/SP level-1 data.
This procedure provides a magnetogram time-series that we interpolate in time to update the lower boundary during the simulation.

The magnetogram time-series together with both large- and small-scale velocity fields prescribe the lower boundary of our model.
This drives our simulation from the bottom by shifting the footpoints of the magnetic field lines, a process often called braiding.
Through this a net upward Poynting flux carries energy into the corona.
Induced currents lead to heating in the corona by Ohmic dissipation, be it through field-line braiding \citep{Parker:1972} or through current sheets formed by coronal tectonics \citep{Priest+al:2002}.

\subsection{Alignment between observations and simulation\label{S:alignment}}

To compare the observations with our simulation results, we need to align the observations spatially, in particular the magnetogram that drives our model and the EIS raster maps which we want to compare with the synthesized coronal emission.
We use the magnetogram in the middle of the time series as a reference and align all magnetograms to it.
By this we can correct for solar rotation as well as for the proper motion of the AR.
To align these photospheric magnetograms with coronal observations, we first align co-temporal snapshots of NFI magnetograms with chromospheric \ion{He}{2} emission recorded by EIS (B-band).
Because \ion{Fe}{12} is recorded on the A-band, we also have to correct for the constant spatial shift between the two EIS bands \citep{Kamio+al:2010a}.
We do this by aligning maps in \ion{Si}{7} (A-band) and \ion{Fe}{8} (B-band), which form at similar temperatures.
As a final step, we align the map in \ion{Fe}{15} recorded by EIS to the X-ray maps taken by XRT with the Ti-poly filter.

Most of the aligned images differ slightly in shape and contrast, and we estimate the alignment in each step to be accurate within about 1.5\unit{arcsec}.
The spatial sampling of EIS and XRT are 1 and 2\unit{arcsec}, respectively.
Because the alignment between any two instruments consists of several steps, and considering the correction for rotation and the AR proper motion, we estimate the overall accuracy to be about 3\unit{arcsec}.
Based on the loop footpoints of the short loop system (SL\,1--3 in \fig{comparison}) we find an alignment residual of 3.5\unit{arcsec} towards north-west.
We subtracted it to make \fig{comparison} clearer.
The circles in \fig{comparison} have a diameter of 3\unit{arcsec}, indicating the accuracy of the alignment.

\section{Hot loops in the core of the active region\label{S:hot.loops}}

Based on the temperatures, densities and velocities in the 3D~MHD model we synthesize profiles of coronal emission lines observable with EIS \citep[following][]{Peter+al:2006}.
In \fig{comparison} we show the observations (left column) and the synthesized maps (right column).
In both observation and synthesized model data we can identify a system of short loops (SL\,1--3) and longer core loops (CL\,1,\,2) in the active region core.

The short loops are identifiable as separate loops in both the observed and synthesized \ion{Fe}{15} emission --- their length, width, and footpoint location coincides within the given accuracy of the alignment.
Loops SL\,1 and 2 are brighter than SL\,3 in both synthetic emission and the observation, so that our model matches the observation of this short loop system very well.
The different curvature of these short loops could indicate a projection effect in the observation due to an inclination of the loops.

For the longer loops there is good agreement between observation and model in the position and shape of CL\,1, even though the synthetic emission is not as strong as the observed emission.
A much weaker (and cooler) loop CL\,2 can only be identified in the model data.
Both CL\,1 and 2 are visible in the simulation only since a few solar minutes.
The temperature of CL\,1 is still rising, which indicates that the heating in the system of field lines around CL\,1 is getting stronger.
Therefore, we expect CL\,1 to develop into a brighter and broader structure in time. This should improve the similarity of the synthesized emission to the observation.

From this comparison of the observed and synthesized emission we can conclude that the model has a distribution of the energy input in space and time to create the actually observed structures.
We now turn to the flows resulting from the dynamics in the active region.

\graphside{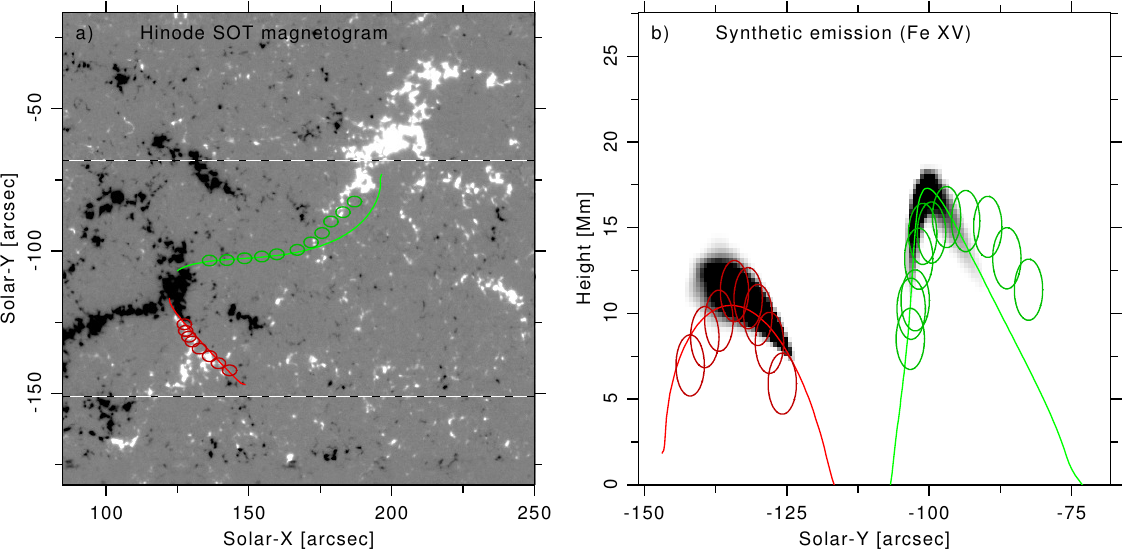}{Stereo}{
Direct comparison of the model intensity structures with the STEREO 3D~reconstruction.
Panel \pan{a} displays the magnetogram (saturated at \eqi{\pm 300\unit{G}}) at the bottom boundary.
Overplotted are the projections of those field lines from the model that cross the maximum of the synthesized emission of the respective loop in the 3D~computational domain.
The circles show the projection of the loops reconstructed from the STEREO observations.
The diameter of the circles indicates the uncertainty in the reconstruction.
In panel \pan{b} we show the synthesized model intensity in \ion{Fe}{15} (284\unit{\AA}) as seen from solar east, i.e., along solar-X, again with the 3D~reconstruction.
The colors denote the core loop (CL\,1, green) and the short loop (SL\,1, red) as introduced in \fig{comparison}, see \sect{hot.loops}.
The dashed white lines in panel \pan{a} indicate the range of solar-Y displayed in panel \pan{b}.}

To investigate the dynamics in the AR, we compare the observed line-of-sight integrated Doppler line-shifts with the synthetic ones (bottom row in \fig{comparison}).
Here we use the \ion{Fe}{12} line, because in the observation this line provides a clearer, less noisy Doppler map than does \ion{Fe}{15}.
The synthesized Doppler maps in \ion{Fe}{12} and \ion{Fe}{15} are quite similar, however.
In general, the synthetic Doppler shifts along the hot loops in the corona correspond well to their observed counterparts (\fig{comparison}c,d).
The northern footpoints of the short loops SL\,1 and 2 are located in a region with magnetic cancellation (see \fig{overview}b) and so experience increased heating.
Here we find upflows (blue-shifts).
The resulting siphon-flows along SL\,1 and 2 towards the southern footpoints are driven by the asymmetric heating.
Rooted farther away from the flux cancellation region we also see a cooler loop (SL\,3) with material draining all along the loop as a result of cooling.

In the AR core we see downflows in both loop legs of CL\,1 together with a rising loop-top.
This is consistent with an emerging loop, where plasma is pushed up (blueshift at apex) and then falls down the legs (redshift at footpoints).
The synthetic Doppler shift at the loop-top corresponds to a vertical velocity of about 2\unit{km/s}, which was also deduced from observations of young loops in an emerging AR \citep[e.g.,][]{Solanki+al:2003}.

\section{STEREO 3D~reconstruction\label{S:reconstruction}}

Besides the Hinode observations, the investigated AR was also observed simultaneously by the STEREO satellites, which allows us to reconstruct the 3D~shape of the coronal loops.
At the observation time, the two satellites had a viewing angle of \eqi{40^{\circ}} between them.
We traced the \ion{Fe}{15} intensity structures observed in the \eqi{284\unit{\AA}} channel of STEREO A and B by first locating both loop legs, then the loop-top in the middle, and finally determining co-spatial points inbetween.
We used the function 'scc\_measure' version 1.15 availabe in the SolarSoft library\footnote{\url{http://www.lmsal.com/solarsoft/}}.
The reconstruction is accurate to several pixels, corresponding to the width of the traced structures that we estimate to about 5\unit{arcsec}, resulting in uncertainties of several Mm.

A comparison of the loop trajectories reconstructed from the STEREO observations with the loops synthesized from the model is shown in \fig{Stereo}.
For this we plot the magnetic field line that passes through the point of maximum emission of the synthesized loop in the 3D~domain for one of the core loops (CL\,1) and one of the short loops (SL\,1).
In the left panel of \fig{Stereo} we show the view from the top (along with the magnetogram at the bottom boundary as the background), and in the right panel we plot the projection of the field lines when looking at the computational domain from the side, now together with the synthesized emission of the loops integrated horizontally through the box (along solar-X).
The reconstruction of the trajectories of the loops observed by STEREO are overplotted as a sequence of circles, the diameter of the circles indicating the uncertainty of reconstructed position.
We find that the synthetic emission from loops CL\,1 and SL\,1 are located within the 3D~reconstruction, both in the horizontal and in the vertical direction.
Most importantly, the reconstructed loops also reach similar heights as the synthesized model loops.
This implies that the model (in general) reproduces the observations also in its 3D~structure.
In the observation we still find a slight inclination of the SL\,1 loop that is not found in the model.

\section{Conclusions\label{S:conclusions}}

We have presented a 3D~MHD model of the corona that is driven by observations of the solar photosphere.
The synthesized profiles of coronal emission lines show strong similarities to the actual observations of the \emph{same} region on the Sun.
This applies to the line intensities and the Doppler shifts that reflect the dynamics within the coronal loops.
Even the spatial distribution of the synthetic emission within the 3D~computational domain occurs roughly at the same location as reconstructed from stereoscopic observations.

In our model all coronal loops examined are heated \emph{predominantly} by Ohmic heating, which is induced by the braiding of field lines through the (horizontal) photospheric motions.
The average Poynting flux into our model corona roughly matches the predicted value of about \eqi{300\unit{W/m^2}}.
Other processes, such as viscous heating of material draining from the corona, also play a role, albeit not the dominant one.
Because the hot structures in the simulation develop at the same locations found in observations, we conclude that the heat in the simulation is deposited in the same places as on the real Sun.
The (asymmetric) heating of the loops and the rise of magnetic field lines leads to flows in the loops that are, again, similar to those found in the observation.

This good match is found even though we cannot resolve the small scales on which the actual dissipation of magnetic energy occurs.
At least the energy deposition at scales accessible to our model, as well as to current coronal observations (above about 500\unit{km}), is well represented by our proxy for the Ohmic heating \eqi{\mu_0 \eta j^2}.
Certainly, on smaller scales many other processes will operate, and we conclude that a heating proportional to the square of the currents provides a good proxy for these sub-grid processes.

The substantial match to the observation shows that the field-line braiding originally proposed by \cite{Parker:1972} provides sufficient energy with the proper distribution in space and time to reproduce characteristic features such as hot coronal loops and their dynamics in an active region.


\begin{acknowledgements}
This work was supported by the International Max-Planck Research School (IMPRS) on Solar System Physics and was partially funded by the Max-Planck-Princeton Center for Plasma Physics (MPPC).
The results of this research have been achieved using the PRACE Research Infrastructure resource \emph{Curie} based in France at TGCC, as well as \emph{JuRoPA} hosted by the J{\"u}lich Supercomputing Centre in Germany.
Preparatory work was executed at the Kiepenheuer-Institut f{\"u}r Sonnenphysik in Freiburg, as well as on the bwGRiD facility located at the Universit{\"a}t Freiburg, Germany.
We thank Suguru Kamio for his help finding active region observations, as well as Li Feng and Navdeep Kaur Panesar for introducing us to stereoscopic 3D~reconstruction, and Neda Dadashi for her suggestions regarding the absolute calibration of the Doppler shifts.
Hinode is a Japanese mission developed, launched, and operated by ISAS/JAXA, in partnership with NAOJ, NASA, and STFC (UK). Additional operational support is provided by ESA and NSC (Norway).
The STEREO/SECCHI data used here were produced by an international consortium of the Naval Research Laboratory (USA), Lockheed Martin Solar and Astrophysics Lab (USA), NASA Goddard Space Flight Center (USA), Rutherford Appleton Laboratory (UK), University of Birmingham (UK), Max-Planck-Institut f\"ur Sonnensystemforschung (Germany), Centre Spatiale de Li\`ege (Belgium), Institut d'Optique Th\'eorique et Appliqu\'ee (France), and Institut d'Astrophysique Spatiale (France).
\end{acknowledgements}

\bibliography{Literatur}
\bibliographystyle{aa}

\end{document}